# 3D Printed Waveguide for Augmented Reality

**Dechuan Sun[1,*], Gregory Tanyi[1], Alan Lee[1], Chris French[2], Younger Liang[3], Christina Lim[1], Ranjith R Unnithan[1,*]**

[1]Department of Electrical and Electronic Engineering, The University of Melbourne, Melbourne, Victoria, Australia

[2]Neural Dynamics Laboratory, Department of Medicine, The University of Melbourne, Melbourne, Victoria, Australia

[3]AR Division, KDH Design Taiwan, Neihu Dist, Taipei City 114, Taiwan

**\* Correspondence:**
Dechuan Sun (dechuan.sun@unimelb.edu.au)

Ranjith R Unnithan (r.ranjith@unimelb.edu.au)

## Abstract

Mass production of augmented reality (AR) waveguides has been challenging due to the intricate nature of the fabrication technique and the high precision required for its optical characteristics. In this paper, we have presented a novel and low-cost approach for fabricating geometric optical waveguides designed for AR applications utilizing 3D printing techniques. To strike a balance between optical performance and fabrication feasibility, we have optimized the conventional geometric waveguide design to facilitate easier fabrication. It is worth noting that our proposed method does not require molding, dicing, and post-surface polishing after printing. A prototype based on this method has been successfully fabricated, showing the immersion between the virtual image and the real-world scene. The proposed method has great potential for adaptation to mass production in various AR applications.

## Introduction

Augmented reality (AR) is a technology that integrates and overlays virtual information onto the user's physical world in real-time, typically achieved through optical see-through near-eye displays. With a diverse range of applications in industry, driving, and education, AR has the potential to become the next-generation computing platform [1-4].

One of the key components in near-eye displays for AR applications is an optical waveguide or combiner. The commercially available optical combiners can be broadly categorized into two major types based on their working principle: geometric and diffractive [2, 5-6]. The geometric waveguide structure was first proposed by Lumus and has undergone continuous optimization over the past twenty years [5]. In this approach, incident light first enters the waveguide through a reflective surface. After undergoing multiple rounds of total internal reflection, the light encounters a "semi-transmissive semi-reflective" mirror array, which couples the light out. On the other hand, the diffractive waveguide structure utilizes diffractive optical elements such as surface relief gratings or volume holographic gratings to couple the light into and out of the waveguide [5-7]. Similar to the geometric waveguide structure, the use of surface relief gratings also provides an effective approach. This involves introducing an incident grating to couple light into the waveguide and substituting the mirror array with an exit grating. Companies such as Microsoft, Vuzix, Magic Leap, and Waveoptics have commercialized devices using this technique [5]. Another diffractive waveguide structure is based on the utilization of volume holographic gratings [5,7], employing a different grating fabrication technique closely related to hologram production [8-10]. While products based on this technology are currently limited in the market, prototypes showcasing the potential of this approach have been demonstrated by Digilens and Sony [5, 7]. A summary of these AR glasses is shown in Table 1. In addition to these approaches, numerous alternative optical waveguides using different methodologies are reported in the literature, such as using holographic optical elements, pinlight configurations, meta surfaces, and polarization devices [11-15]. However, these technologies have not yet been commercialized due to various challenges in the fabrication process.

Both geometric and diffractive waveguides exhibit advantages and disadvantages. Diffractive waveguide technology relies on the deposition of a thin film on glass substrates, eliminating the need for glass slicing and bonding processes used in fabricating geometric waveguides. This provides advantages in terms of flexibility in design and production. However, a notable drawback is the presence of chromatic aberration or the appearance of the rainbow effect [16]. Moreover, diffractive waveguides may have limitations in their field of view. These issues primarily stem from the inherent selectivity of diffraction gratings with respect to light input angles and wavelengths [16]. Additionally, the refractive index of the waveguide also plays a role in these limitations [17]. In contrast, the geometric waveguide structure can offer a large field of view, excellent image quality, and no colour dispersion. However, the commercialization of both methodologies faces challenges due to limitations in mass production capabilities [16-18]. Consequently, most AR glasses on the market are expensive.

Recently, ultra-clear transparent UV resin, combined with 3D printing technology, has facilitated the production of 3D-printed optics that exhibit a quality comparable to that of traditional optical components. [19-23]. This specialized resin is formulated to minimize cloudiness and demonstrate low shrinkage after UV light curing, thereby enabling the fabrication of AR optical combiners that offer exceptional clarity and transparency. However, the application of 3D printing in the AR field has received limited attention. To the best of our knowledge, 3D-printed waveguides have not been reported in previous literature. In this paper, we have optimized the conventional geometric waveguide structure with consideration for fabrication and developed a cost-effective 3D printing platform for fabricating the waveguide that integrates three dielectric reflectors through our patented manufacturing pipelines [24].

## Simulation results

We have optimized our geometric waveguide design using the finite element method implemented in COMSOL Multiphysics to strike a balance between optical performance and fabrication feasibility. For example, if the thickness of the waveguide is too thin, it may result in breakage when removing the printed sample from the glass printing bed. The dimensions of our geometric AR waveguide are 31mm x 26mm x 7.6mm. A 3D-printed triangular prism is utilized to facilitate the efficient coupling of light into the waveguide. The prism is placed at an angle of 50 degrees relative to the bottom surface of the waveguide. This particular angle ensures total internal reflection within the waveguide, enabling optimal light transmission along its path. For simplicity, the design incorporates three dielectric reflectors, integrated at an angle of 25 degrees relative to the bottom surface, as illustrated in Fig. 1. These reflectors serve the purpose of directing the light toward the observer's eye. The projector used in our design is a general LCOS projector with a 40° field of view, which emits a virtual image that is subsequently coupled into the waveguide through the triangular prism. Then, we conducted simulations using COMSOL to evaluate the effectiveness of the proposed design. The simulation parameters included a refractive index of 1.53 for the resin material used in the waveguide. Additionally, considering the propagation loss, the dielectric reflectors employed in the waveguide were assigned a transmittance ratio of 90% based on actual measurements, and wall boundary conditions were applied. Furthermore, we used an ideal lens with a focal length of 25mm and applied optimal lens boundary conditions to simulate the human eye. Fig. 1 shows the simulated ray tracing results, illustrating the propagation of light rays through the waveguide, their convergence at the retina, and the formation of the final image. An example of the computationally reconstructed University of Melbourne's crest is also shown in the figure. To achieve this, an object (bitmap file of the crest) was loaded into the model, where the spatial distribution of rays released from the object's surface was proportional to the bitmap's values. With the release of 23000 rays, we achieved a magnified and perfectly stitched image at the image plane, achieving a magnification ratio of 1.4. This waveguide design not only demonstrates its efficacy in providing high-quality image reconstruction but also takes into consideration its feasibility for fabrication.

## Manufacturing process

The optimized geometric AR waveguide was fabricated using a semi-custom level 3D printer. Our platform is constructed based on the Phrozen Sonic mini 8K resin 3D printer, which is a custom-level liquid-crystal display (LCD) 3D printer. Briefly, a printing bed is immersed in a transparent resin vat and exposed to UV light emitted by an array of LEDs at the bottom. The UV light passes through an LCD screen, acting as a masking element. This reveals the image pattern on the printing bed and selectively cures corresponding pixels. This layering process is repeated for subsequent layers. The LCD screen provides an impressive ultra-high pixel resolution of 22μm, and the dual linear rails offer a stable z-axis step size of 10 μm.

To enhance the uniformity of light distribution, a diffuser and an absorber were placed in front of the light source. Additionally, another diffuser was placed on top of the LCD screen to address the issue of LCD inter-pixel gaps and ensure a seamless display of the image pattern (Fig. 2). In practice, we found that the second diffuser did not significantly impact the x-y plane resolution, particularly when the z-axis step size was under 20μm. This observation held true, especially for objects lacking intricate details. The resin vat incorporated a thin perfluoroalkoxy (PFA) film as a release liner, aiming for a non-adhesive surface that facilitated the removal of the printed objects upon completion. To enhance non-stick properties, a thin layer of polytetrafluoroethylene (PTFE) was coated on the PFA film. We used NOVA3D ultra-clear resin, which had a relatively stable refractive index of around 1.53 in the visible light wavelength range. At 25°C, it exhibited a volume shrinkage of 3.6% upon curing. Unlike professional inkjet printers, which could achieve smooth surface finishes without additional post-processing, LCD printers could not offer comparable surface finish roughness. A strategy to mitigate this issue was to print the top and bottom waveguide components separately on glass printing beds, followed by integrating them with dielectric reflectors (as seen in Fig. 2). This approach improved the overall surface finish of the waveguide and potentially enhanced its performance. Regarding the prism's sloped surface, achieving a smooth finish is not essential since it could be glued to the projector at a later stage.

We first adhered plain glass slides to the metal printing bed. Using a z-axis step size of 10μm, the top and bottom components of the waveguide were printed with the flat surface facing down towards the glass slides. To improve the printing quality, antialiasing was implemented on the image pattern during the printing process. This technique effectively blurred the image pattern, reducing the presence of jagged edges and thereby enhancing the subpixel resolution of the final product. The total printing time was around 2 hours, and we did not add any supports to the model. Once the printing process was finished, the printed parts, along with the glass slides, were removed from the printing bed and immersed in a shallow resin tank. The sample was examined under a microscope to assess its quality. Any debris on the surface could be removed by using nitrogen to blow it away gently. It was crucial to avoid using isopropyl alcohol during this cleaning process, as it could potentially diminish the level of transparency. Dielectric reflectors were then carefully placed into the corresponding slots within the resin, effectively preventing bubble generation during the placement. Subsequently, the combined sample was taken out and cured within a nitrogen chamber under UV, with proper pressure applied on the top glass. Then the sample was transferred to a 40°C hot plate and allowed to undergo a 30-minute thermal treatment to relieve the sample's internal stress that accumulated during the curing phase [25]. Finally, the glass slides were removed using a sharp blade.

## Experimental results

The integration of diffusers led to a significant enhancement in printing quality. The use of the first diffuser, placed in front of the light source, resulted in substantial improvements in light uniformity. The incorporation of the second diffuser on top of the LCD screen further enhanced the uniformity of illumination, effectively mitigating the LCD inter-pixel gaps issue. Without this diffuser, the printed sample displayed many fine lines within its structure (Fig. 3).

In the design, we utilized dielectric thin-film interference filters as reflectors. The surface roughness of the reflector was evaluated using atomic force microscopy (AFM), and the corresponding results are presented in Fig. 4. According to standard surface roughness definitions, the measured RMS roughness of the reflector is $R_q = 1.40$ nm, and the mean roughness is $R_a = 1.09$ nm. In addition to the surface roughness measurements, we assessed the transmittance ratio of the reflector at various angles within the visible light wavelength range. The reflector demonstrates a transmittance ratio of approximately 90% at both 0-degree and 65-degree angles, corresponding to real-world light transmission scenarios (Fig. 4).

The combined printed waveguide has a thickness of 3mm. The planar surface printed on the glass slide exhibits good roughness characteristics. It has an RMS roughness of 1.49 nm and a mean roughness of 1.38 nm (Fig. 4). Regarding the transmittance ratio, the waveguide exhibits low transmittance from 400 nm to 420 nm. However, beyond 420 nm and up to 700 nm, the transmittance ratio consistently remains high, hovering around 90%, which renders it well-suited for a wide range of optical applications within this specific spectral region (Fig. 4).

To validate the efficacy of the design, a prototype of the printed AR waveguide, incorporating three dielectric reflectors, was successfully fabricated, as depicted in Fig. 5a-b. Moreover, the AR waveguide was tested using a commercial microprojector. Fig. 5c-d display an example, showing the projected image and the image captured through the waveguide. However, one limitation of the current waveguide is the colour fidelity of black, which is influenced by the brightness of the projector. Further optimization of the projector is also necessary. A video showing the overlapping can be found in the supporting information.

## Conclusions

In this study, we present an innovative approach for the fabrication of geometric AR optical combiners intended for applications in augmented reality. Instead of depending on high-cost inkjet 3D printers, renowned for their exceptional surface smoothness, we have developed a strategy focused on utilizing a cost-effective LCD 3D printer. This printer is custom-adapted to achieve significantly improved surface roughness without the need for molding, dicing, and post-polishing. On the other hand, the inkjet printer requires resin with an extremely low viscosity, which limits the choice of available photoinitiators, stabilizers, and UV absorbers. This constraint, in turn, can raise concerns about resin yellowing. In contrast, the LCD printer offers a broader range of options for resin selection. In the current design, we have integrated only three dielectric reflectors for simplicity. However, additional reflectors can be integrated to expand the eyebox and field of view. Finally, we have successfully fabricated a prototype of the proposed optical combiner, creating a virtual image that seamlessly overlaps with the real world. This achievement demonstrates the system's potential for low-cost mass production.

## Supporting Information

"supplementary video.mov", showing a video captured using the printed AR waveguide.

## Acknowledgments

This work was supported by Research Funding from Jarvish.

## Author Contributions

## Conflict of interest

## Data availability statement

## Table

| Waveguide | Company | Cost |
|---|---|---|
| Geometric | Our design | waveguide: $5 |
| | Lumus | $3000 |
| Surface relief gratings | Microsoft Hololens 2 | $3500 |
| | Vuzix M4000 | $2500 |
| | Magic Leap 2 | $3299 |
| | Waveoptics Katana | Currently not available |
| Volume holographic gratings | Digilens | Currently not available |
| | Sony | Currently not available |

**Table 1. AR glasses available on the market.** Due to the challenges in manufacturing AR waveguides, most AR glasses require a long lead time. Our design features a highly affordable waveguide, priced at around $5. However, due to a non-disclosure agreement, we are unable to provide the price of the projector.

# Figures

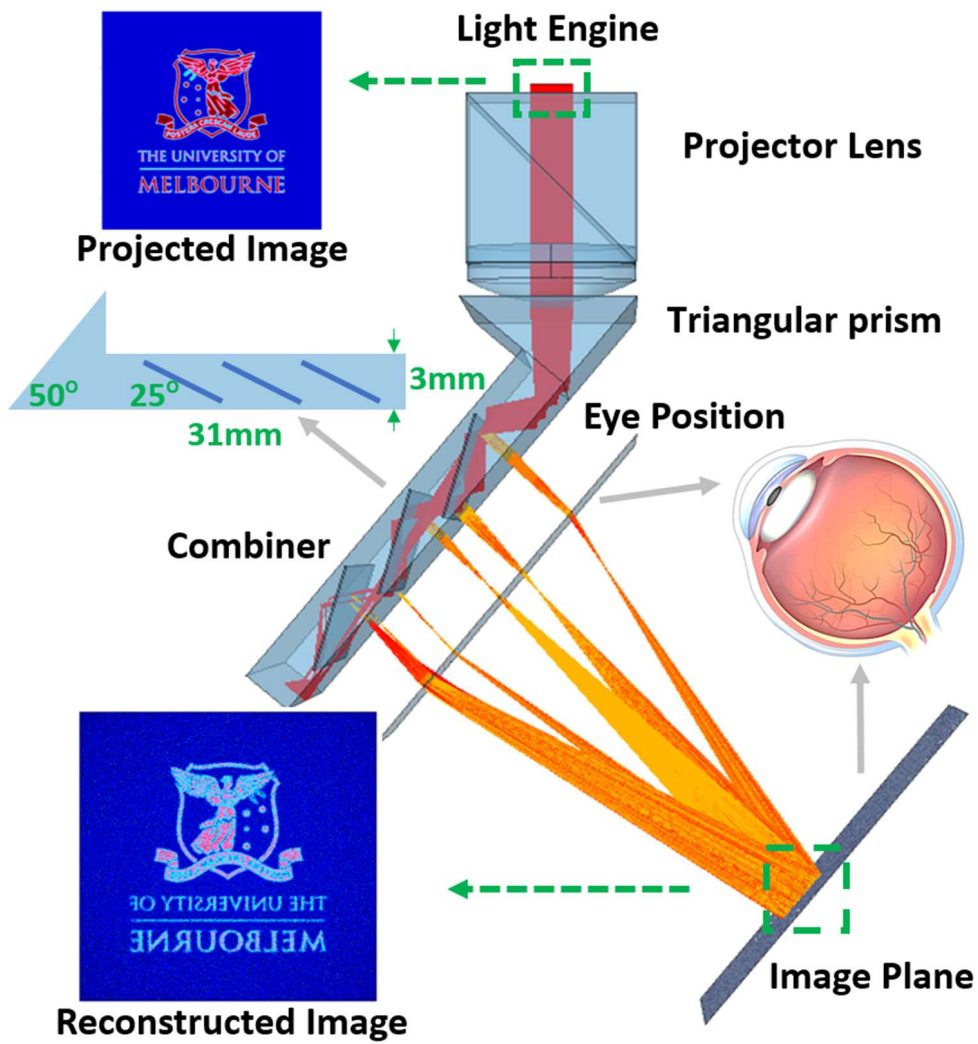

**Figure 1**. **Ray trace showing light path, image combination, and pupil expansion as light propagates from coupling optics to the user's eye.** The object (bitmap file) is loaded into the model in such a way that the spatial density of rays released from the object surface is proportional to the values in the imported bitmap. The reconstructed image at the image plane, generated using 23,000 rays, displays a magnified, perfectly stitched, and flipped image. A 2D schematic of the waveguide illustrates the critical angles and dimensions of the combiner.

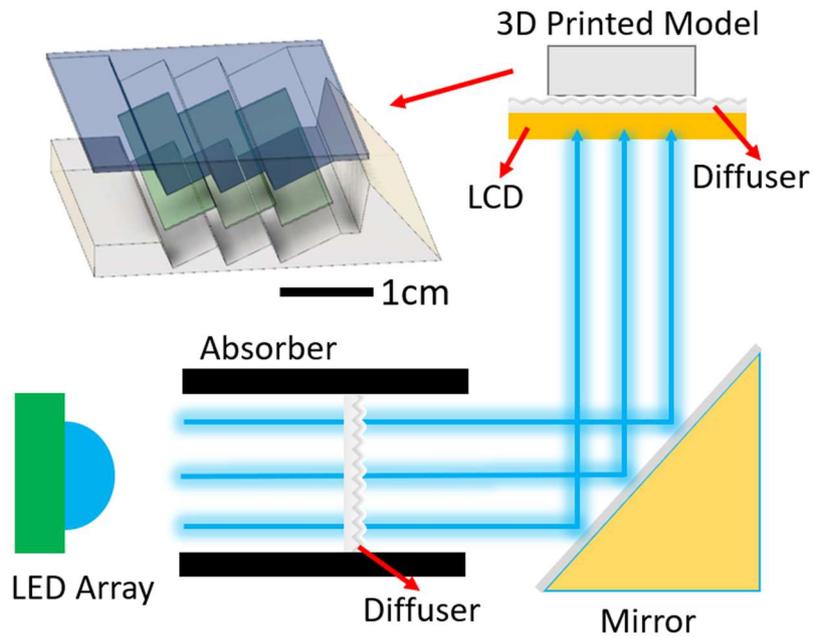

**Figure 2. The adapted printing platform and the fabrication process.** Schematics illustrate the modified 3D printing platform. To enhance light distribution uniformity, a diffuser and an absorber were placed in front of the light source. Additionally, another diffuser was positioned on top of the LCD screen. The top and bottom components of the waveguide were printed separately on glass printing beds and later integrated with the dielectric reflectors.

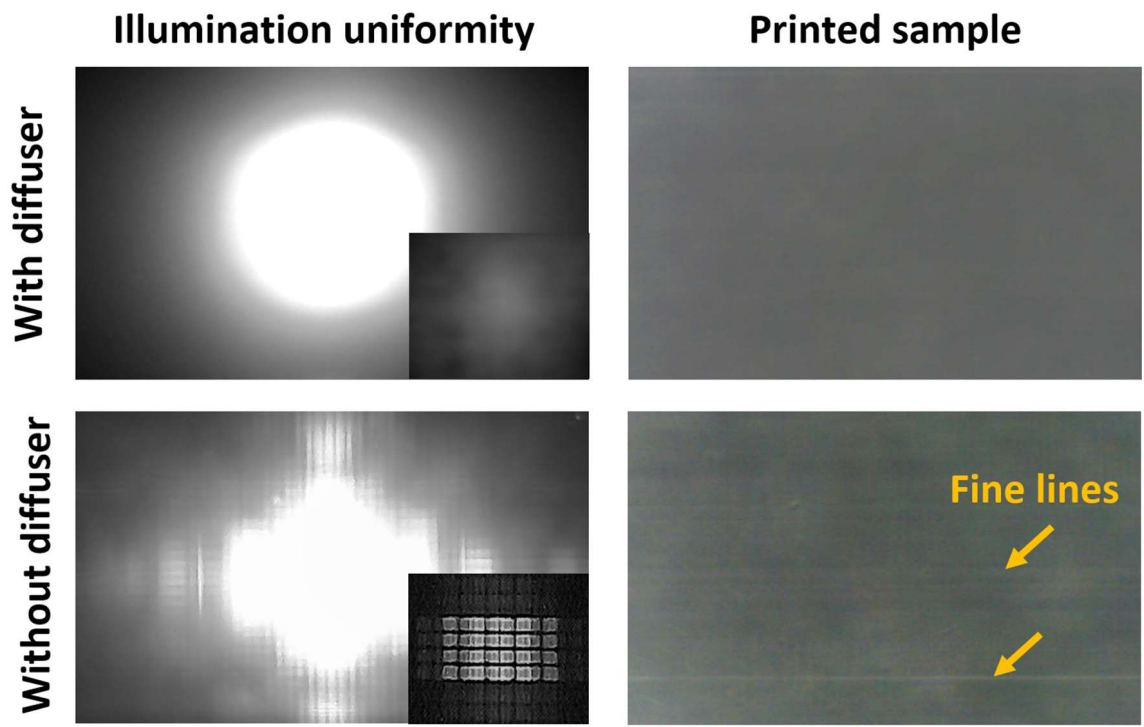

**Figure 3. Illumination uniformity and LCD inter-pixel gap issues. (Left)** A comparative example showing the illumination uniformity with and without the use of the light source diffuser. As the intensity of the light source is too strong, saturating the camera sensor, an optical filter is employed to reduce the LED intensity, revealing illumination patterns depicted in the bottom right subfigures. **(Right)** Without the LCD diffuser, the printed sample displayed numerous fine lines within the body, as indicated by the arrows. However, these lines disappeared after the diffuser was applied. To enhance line visibility, the transparent sample was placed on a piece of black paper and observed under a microscope.

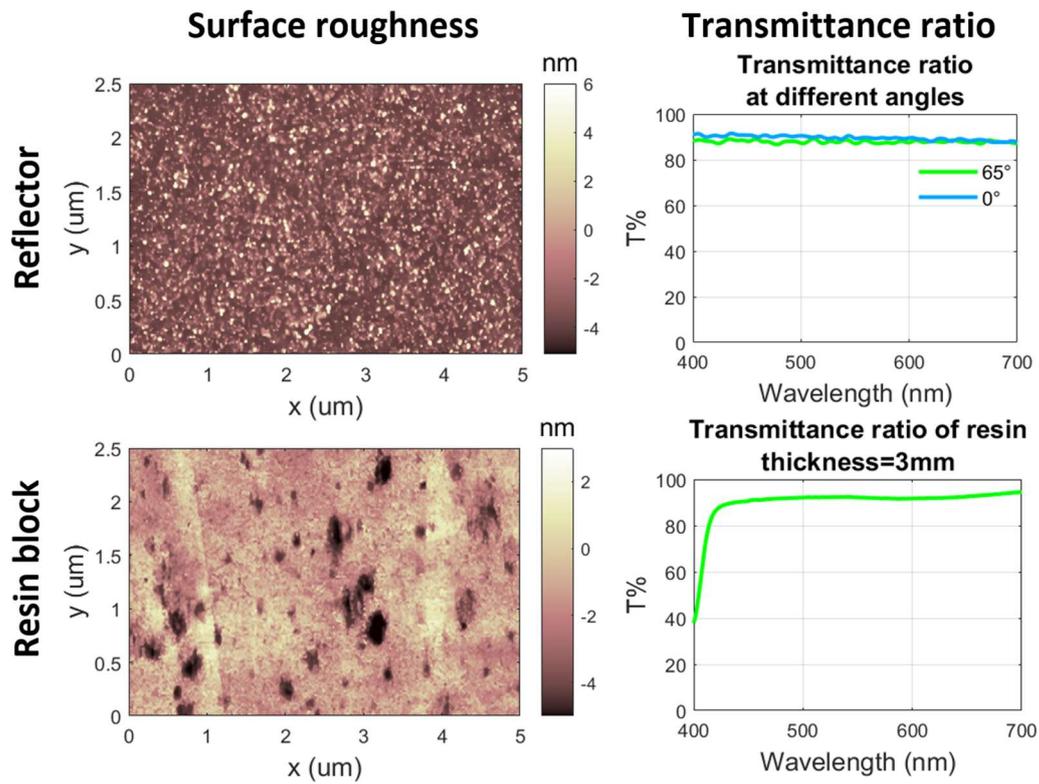

**Figure 4. Surface roughness and transmittance ratio of the reflector and the 3D printed sample with a thickness of 3mm. (Left)** The 2D surface roughness maps of the reflector and a printed resin block were measured over a small area (5 μm × 2.5 μm). **(Right)** The transmittance ratio of the reflector was measured at an incident angle of 0-degree and 65-degree. The transmittance ratio of the 3D printed resin block was measured at an incident angle of 0-degree, showing an average transmittance ratio of 92.3% above 420nm.

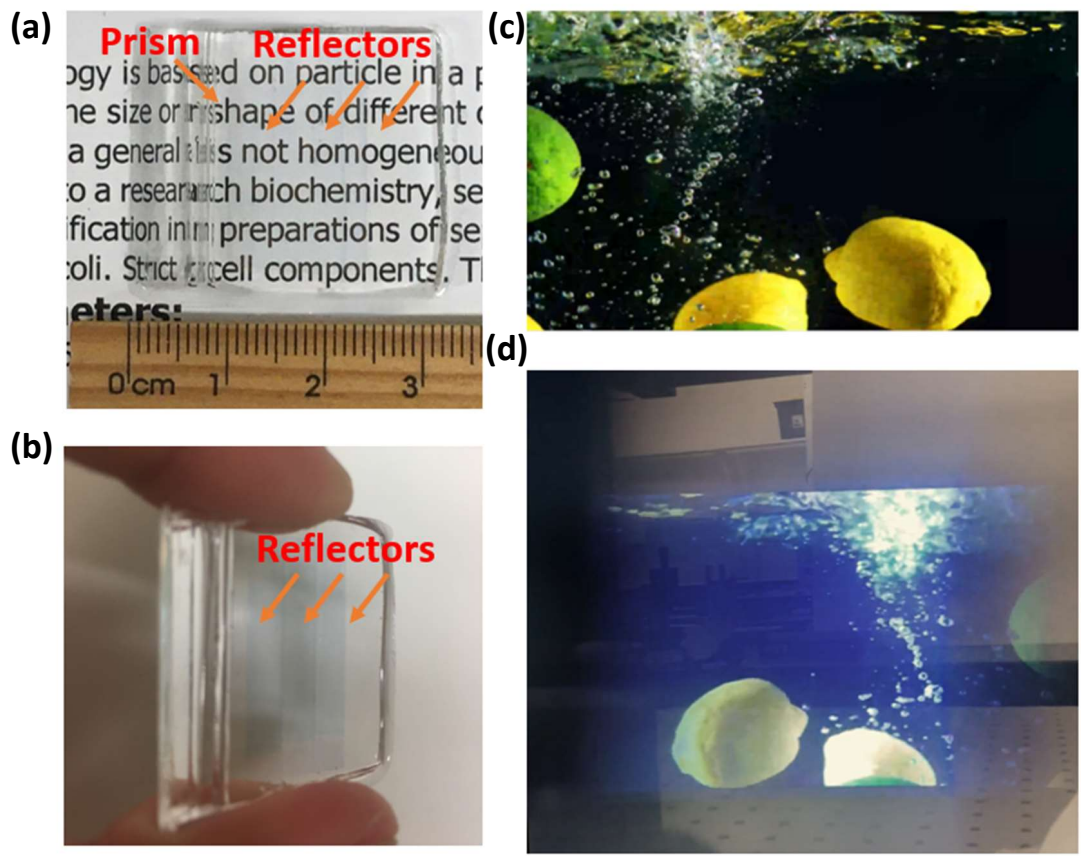

**Figure 5. Experimental results.** (a) A prototype of the printed waveguide that integrates three reflectors. (b) A prototype is placed at an angle to show the reflectors. (c) The projected image. (d) The overlap of the virtual image with the real world. The virtual image is horizontally mirrored due to the combiner design.